\documentclass[12pt]{iopart}

\expandafter\let\csname equation*\endcsname\relax
\expandafter\let\csname endequation*\endcsname\relax

\usepackage{amssymb}
\usepackage{graphicx}
\usepackage[pdftex,bookmarks,colorlinks,breaklinks]{hyperref}
\hypersetup{linkcolor=red,citecolor=blue,filecolor=dullmagenta,urlcolor=blue}
\usepackage{amsmath}
\usepackage{mathrsfs}
\usepackage{amsfonts}

\usepackage[utf8]{inputenc}
\usepackage[T1]{fontenc}
\usepackage{bbm}
\usepackage{color}
\usepackage{braket}
\usepackage{soul,xcolor}

\newcommand{\rgoe}{\langle\tilde{r}_\lambda\rangle_{\mathrm{GOE}}}

\newcommand{\rhf}{\langle\tilde{r}_\lambda\rangle_{\mathrm{HF}}}

\newcommand{\roe}{\langle\tilde{r}_\lambda\rangle_{\mathrm{OE}}}
\newcommand{\Ne}{{N_{\rm s}}}

\usepackage{orcidlink}
\usepackage{fancyhdr}
\usepackage{calc}
\begin{document}

\title{\textsf{\textbf{Onset of universality in the dynamical\newline mixing of a pure state}}}

\author{M Carrera-N\'{u}ñez$^{1,2}$\,\orcidlink{0000-0002-1526-4864},\,%
A M Mart\'inez-Arg\"uello$^{3}$\,\orcidlink{0000-0002-6422-0673},\,\newline%
J M Torres$^{3}$\,\orcidlink{0000-0002-7947-6962}\, and\,\,%
E J Torres-Herrera$^{3}$\,\orcidlink{0000-0001-7619-5105}%
}
\address{$^{1}$Departamento de Ciencias Naturales y Exactas, Universidad de %
Guadalajara, Carretera Guadalajara---Ameca Km. 45.5 Ameca, %
Jalisco 46600, Mexico}

\address{$^{2}$Departamento de F\'isica, Universidad Aut\'onoma Metropolitana-Iztapalapa, Apartado Postal 55-534, Ciudad de M\'exico 09340, Mexico}

\address{$^{3}$Instituto de F\'isica, Benem\'erita Universidad Aut\'onoma de Puebla, Apartado Postal J-48, Pue., 72570 Puebla, Mexico}
\ead{\textcolor{blue}{blitzkriegheinkel@gmail.com}}


\begin{abstract}

\noindent We study the time dynamics of random density matrices generated by evolving the same pure state using a Gaussian orthogonal ensemble (GOE) of Hamiltonians. We show that the spectral statistics of the resulting mixed state is well described by random matrix theory (RMT) and undergoes a crossover from the GOE to the Gaussian unitary ensemble (GUE) for short and large times respectively. Using a semi-analytical treatment relying on a power series of the density matrix as a function of time, we find that the crossover occurs in a characteristic time that scales as the inverse of 
the Hilbert space dimension.
The RMT results are contrasted with a paradigmatic model of many-body localization in the chaotic regime, where the  GUE statistics is reached at large times, while for short times the statistics strongly depends on the peculiarity of the considered subspace.
\end{abstract}

%
\vspace{2pc}

\pagestyle{fancy}
\fancyhf{}
\fancyhead[L]{Onset of universality in the dynamical mixing of a pure state}
\fancyhead[R]{\thepage}

\section{Introduction}
\label{sec:introduction}

Random quantum states appear quite naturally in many problems of quantum mechanics. In quantum information theory, for instance, they are indispensable to study the average degree of entanglement in systems where no analytical solution is available~\cite{Wootters1990,Collins2016}. Applications are also found in the theory of open quantum systems, where the effects of a noisy interaction invariably lead to a dynamically generated random state~\cite{Zyczkowski2011,Pineda2015}. It is in this context that interesting recent  developments have been achieved, where attention has been given to characterize the ensemble of mixed random states inspired in the results of random matrix theory (RMT)~\cite{Pineda2015,Chamon2014,Yang2017,Sarkar2021,Vinayak2012}. For instance, the eigenvalues of mixed random states belonging to the Hilbert-Schmidt ensemble follow a Marchenko-Pastur distribution~\cite{Collins2016,Zyczkowski2011,Sommers2004,Znidaric2007,Bassler2009}.

An interesting phenomenon appearing in RMT is the transition in the spectral statistics from the Gaussian orthogonal ensemble (GOE) to the Gaussian unitary ensemble (GUE) obtained from varying a single parameter in the Hamiltonian~\cite{pandey1983gaussian,Schierenberg2012,Lenz1991,Schweiner2017}. This phenomenon has been experimentally verified~\cite{So1995,schafer2002transition}, and it has also been predicted in physical scenarios involving driven non-integrable quantum systems~\cite{haake1987classical,Mierzejewski2013}. An interesting question is whether this effect is also present in the case of mixed random states. A random density matrix can be obtained as an incoherent mixture of random pure states, that in turn can be obtained by a random evolution of a fixed initial state. For large times, it is known that such a state belongs to a Wishart ensemble and some spectral properties typically correspond to the GUE~\cite{Zyczkowski2011}. This type of GUE statistics has also been observed in the entanglement spectrum or eigenvalues of a reduced density matrix in the context of thermalization of coupled quantum systems~\cite{Mierzejewski2013,Geraedts2016,Regnault2016} and random quantum circuits~\cite{Zhou2020,Iaconis2021}. One can expect, however, that the way such a density matrix is reached as the evolution develops can display interesting features before a universal regime is attained.

In this work we focus on the transient behavior of quantum states before reaching the final random state following the universal GUE statistics. We show that a generic crossover exists for a time scale that depends on the system size. Starting from GOE, the spectral statistics of the state transits in a smooth way to GUE, a transition that is equivalent to that occurring in the spectral statistics of the Hamiltonian when time-reversal symmetry is broken~\cite{Beenakker1997,assmann2016quantum,schweiner2017magnetoexcitons}. Here, however, the transition is not related to the breaking of anti-unitary symmetries in the Hamiltonian, but with the elements of the density matrix, i.e., real or complex uncorrelated Gaussian random variables.  We analyze this behavior using a semi-analytical treatment for the short-time density matrix. Furthermore, we show that the transient behavior to the GUE universal form is also preserved in more realistic situations. For this purpose, we consider that the time evolution of the density matrix is  generated by an ensemble of Hamiltonians describing nearest-neighbor interacting spin-1/2 particles with on-site disorder, in the chaotic regime. In this case, the transition to the GUE statistics is also reached for large enough times, while for short times the statistics shows a strong dependence on the peculiarity of the considered subspace.

The paper is organized as follows. In section~\ref{ensemble}, we explain the dynamical approach that we  use to generate random mixed states.  Section~\ref{onset} is devoted to show that a crossover from a GOE-like to a GUE-like statistics in the time-dependent density matrices is possible during the evolution generated by Hamiltonians  taken from the GOE. A detailed numerical analysis of the mixed state spectral statistics  is given in section~\ref{sec:specstat}. We revisit the problem when the evolution is governed by spin-$1/2$ Hamiltonians in section~\ref{sec:Heisenberg}. We show that a transition in the statistical properties is also present for two different subspaces, one of them with a single excitation and the other with half of the individual spins excited. Conclusions and final remarks are given in section~\ref{conclusions}.


\section{Ensemble of a randomly evolved pure state}
\label{ensemble}

In this section we present a simple model consisting of an ensemble of quantum systems initially prepared in the same  state $\ket{\Psi_0}$ and evolving under the influence of a different Hamiltonian. We assume a finite dimensional system, so that the state of each member of the ensemble can be represented by a state vector in a Hilbert space of dimension $N$. The dynamics of the $l$-th member is governed by a random Hamiltonian $H_l$ belonging to an ensemble that will be later specified. The time-dependent state vector of each
realisation of the ensemble can be expressed as
\begin{equation}
\label{eq:PsiofT}
\ket{\Psi_l(t)}= U_l(t)\ket{\Psi_0}, \quad U_l(t)=e^{-iH_lt},
\end{equation}
where we have chosen units where $\hbar=1$ and we have introduced the time-evolution operator $U_l(t)$ dependent on the Hamiltonian $H_l$. Using the pure-state evolution generated by each Hamiltonian, equation~(\ref{eq:PsiofT}), one can find that the state of the whole sample of $\Ne$ pure states is described  by their incoherent sum  and is given by the following density matrix
\begin{equation}
\label{eq:RT}
\rho(t) = \sum_{l=1}^{\Ne} \frac{1}{\Ne} | \Psi_l(t) \rangle \langle \Psi_l(t) | .
\end{equation}
This density matrix describes a mixed quantum state where one can assign the same probability $1/\Ne$ to each member of the sample $\ket{\Psi_l(t)}$. Constructed in this way it could, for instance, correspond to $\Ne$ particles initially prepared in the same state, but each one subjected to a different Hamiltonian evolution. In general, however, one could think of $\Ne$ identical quantum systems  starting from the same condition, but evolving in different ways.

We will focus our attention on Hamiltonians with real entries in certain basis and initial states with random real probability amplitudes in that same basis. In particular, we will start considering each $H_l$ as a member of the GOE. In such case, $H_{l}\in\mathrm{GOE}$, the time evolution of the particles is ruled by an unknown Hamiltonian over which we only assume that it is time-reversal invariant. In that sense and from the physical point of view, equation~(\ref{eq:PsiofT}) and equation~(\ref{eq:RT}) represent a \emph{generic} time-evolution of quantum particles~\cite{Lutz1999}. For arbitrary values of time,  the mixed state in equation~(\ref{eq:RT}) will be given by a density matrix with complex entries. For large enough values of time, it is plausible to infer that real and imaginary parts will have, on average, the same weight. Invoking the central limit theorem (CLT), it is not hard to realize that the resulting density matrix will share some properties with members of the GUE. However, this is not the case for short times. Indeed if one takes the incoherent sum of real random states, the resulting density matrix will resemble a member of the GOE. Numerical evaluation of spectral quantities demonstrates  this fact as we will see later in this work, even when each $H_l$ is not a member of the GOE but includes certain random components. For now, let us connect this behavior with previous results involving a crossover from the GOE to the GUE statistics in RMT.


\section{GOE to GUE crossover: Onset of universality}
\label{onset}

It is known that a GOE to GUE statistics crossover can be observed by tuning a real parameter $\alpha$ in a Hamiltonian of the form
\begin{equation}
\label{eq:Hgoetogue}
H   =S+i\alpha A
\end{equation}
with $S$ a real symmetric Gaussian matrix, and $A$ a real antisymmetric Gaussian matrix having variances
\begin{equation}\label{crossoverH2}
    \langle (S_{n,n})^2\rangle=        
   2 \langle (S_{n,m\neq n})^2\rangle=
    2\langle (A_{n,m\neq n})^2\rangle  =   1.   
\end{equation}
In previous works it has been observed that, for a given dimension of the system $N$, the transition takes place at $\alpha\sim 1/\sqrt{N}$~\cite{Schierenberg2012,Mehta1983,Kanazawa2020}. Using this result, we will elucidate how a similar crossover arises in time for the simple model of the random density matrix explained above. Furthermore, we will show that at the stage when the GUE statistics is achieved, all other features of a random density matrix are attained as well. It is in this sense that, in this case, the GOE to GUE crossover marks a prelude to  the universal feature of a dynamically generated density matrix that we refer to as onset of universality.


\subsection{Short-time dynamics}
\label{shorttime}

In order to study the aforementioned transition, we consider the short-time behavior of the density matrix by evaluating its expansion as a power series of the time parameter $t$. For this purpose, we first consider the following truncated form of the evolution operator
\begin{equation}
\label{eq:Ut}
U_l(t)=e^{-iH_l t}
\simeq P(H_lt)
= 1-itH_l-\frac{t^2}{2}H_l^2+i\frac{t^3}{6}H_l^3.
\end{equation}
A sufficient condition for this expansion to be valid is that the largest eigenvalue of $H_l$ multiplied by $t$ remains much smaller than one. Here we assume that each $H_l$ is a member of the GOE of size $N$ with largest eigenvalue given by $E_{\rm max}=\sqrt{2N}$ which translates into the following condition in time $t\ll 1/\sqrt{2N}$. An estimation of the truncation error can be obtained by considering the difference operator $D(H_l t)=\exp(-iH_t t)-P(H_lt)$, and
noting that its spectral norm can be bounded as
$|D(E_{\rm max}t)|\le (E_{\rm max}t)^4/4!$.

Using the Taylor expansion of the evolution operator, it is possible to evaluate the density matrix up to any given order in time in the following way
\begin{equation}
\rho(t)=\frac{1}{\Ne}\sum_{l=1}^\Ne U_l(t)\rho_0U_l^\dagger(t)
=
\frac{1}{\Ne}\sum_{l=1}^\Ne\sum_{k=0}^\infty \sigma_{l,k} t^k,
\end{equation}
where $\rho_0=\ket{\Psi_0} \bra{\Psi_0}$. As the GOE is invariant under orthogonal transformations, using a basis state of the Hamiltonian as initial condition is equivalent to choosing a random state with real entries. For this reason, and without loss of generality, we set the initial state to $\ket{\Psi_0}=\ket0$, where $\ket{n}$ is a state of the basis where all the random Hamiltonians are constructed, and $n=0,1,\dots, N-1$. In this way, $\rho_{0}=\ket{0} \bra{0}$ represents the density matrix of the initial pure state. Since we are interested in small values of $t$, we proceed as we did with the expression in equation~(\ref{eq:Ut}) and we keep only the first four terms in the  expansion (third order in time). These terms are given by
\begin{eqnarray}
\label{eq:rhol0}
\sigma_{l,0}&=\rho_0=\ket{0} \bra{0},\quad \sigma_{l,1}=i[\rho_0,H_l],\\
\label{eq:rhol2}
\sigma_{l,2}&=H_l\rho_0H_l-\frac{1}{2}\{H_l^2,\rho_0\},\\
\label{eq:rhol3}
\sigma_{l,3}&=\frac{i}{2}[H_l\rho_0H_l,H_l]+
	   \frac{i}{6}[H^3_l,\rho_0],
\end{eqnarray}
where $[H,H']$ and $\{H,H'\}$ respectively stand for the commutator and anticommutator between $H$ and $H'$. We consider terms up to third order in $t$ as it is at this order where the first antisymmetric full matrix appears in $\sigma_{l,3}$, whenever each Hamiltonian $H_l$ is real. From expressions~(\ref{eq:rhol0})-(\ref{eq:rhol3}), one can note that the only terms represented by full matrices are $H_l\rho_0H_l$, and $[H_l\rho H_l,H_l]$. All other elements contain nonzero entries only in the first column and in the first row. Using this fact, it is possible to separate the terms consisting of full matrices by rewriting the density matrix in the following way
\begin{equation}
\label{rhosplit}
\rho(t)\simeq \tilde\sigma(t)+\sigma(t).
\end{equation}
The first term contains only a nonzero element in the first position of the main diagonal. Other $2N-1$ nonzero elements are present in the first column and first row of the matrix, as evidenced by the expression
\begin{equation}
\label{sigmatilde}
\tilde\sigma(t)=
\ket{0}\bra{0}+\sum_{n\neq 0}
\left(
a_n(t)\ket{0}\bra{n}+a_n^\ast(t)\ket{n}\bra{0}
\right)+\frac{t^2}{2}\mathbbm{1},
\end{equation}
where $\mathbbm{1}$ is the identity matrix and where we have introduced the matrix element
\begin{equation}
a_n(t)=\left[\sum_{l=1}^\Ne\frac{1}{\Ne}\left(
-iH_lt-\frac{1}{2}H_l^2t^2+\frac{i}{6}H_l^3t^3
\right)\right]_{n,0}.
\end{equation}
The second term in equation~(\ref{rhosplit}) is a full matrix, therefore containing $N^2$ nonzero elements that will dominate the statistics for large $N$. This term has the following form
\begin{eqnarray}
\label{crossoverrho}
\sigma(t)=&
\frac{t^2}{\sqrt{2\Ne}}\left(B+it\frac{\sqrt N}{2}D\right),
\end{eqnarray}
where we have chosen to write it in terms of the real symmetric matrix $B$ and the real antisymmetric matrix $D$ with zero mean value and matrix elements given by
\begin{eqnarray}
B_{n,m}&=\sqrt{\frac{2}{\Ne}}\sum_{l=1}^\Ne
[H_l]_{n,0}[H_l]_{0,m}-\sqrt{\frac{\Ne}{2}}\label{matricesrhoa}\delta_{n,m} , \\
D_{n,m}&=\frac{\sqrt{2}}{\Ne}
\sum_{l=1}^\Ne
\left(
[H_l^2]_{n,0} [H_l]_{0,m}
-
[H_l]_{n,0}[H_l^2]_{0,m}
\right).\label{matricesrhob}
\end{eqnarray}
From the CLT and taking into account the properties of $H_l$, it is not hard to find that the variances of $B$ and $D$ are given by
\begin{equation}\label{sigmarho}
\langle (B_{n,n})^2\rangle=        
2 \langle (B_{n,m\neq n})^2\rangle=
2\langle (D_{n,m\neq n})^2\rangle  =   1.   
\end{equation}
We have considered $H_l$ as an element of the GOE, however, this treatment is also applicable to random symmetric Hamiltonians with zero mean and variances as $S$ in equation~(\ref{crossoverH2}).

From the discussion after equation~(\ref{eq:Ut}), we found the condition $t\ll 1/\sqrt{2N}$ for the validity of the truncation of the evolution operator. Therefore, our expressions in this section have to be restricted to this limit in time. The transition to GUE statistics can be faithfully explained from equation~(\ref{crossoverrho}) taking into account the aforementioned time restriction. According to the discussion in the previous subsection, this transition should manifest itself in time when the imaginary part of the density matrix is $1/\sqrt N$ times smaller than the real part. In equation~(\ref{crossoverrho}), this corresponds to a time given by 
\begin{equation}
\label{eq:guetime}
t\sim 2/N.
\end{equation}
For large enough $N$ the condition $t\ll 1/\sqrt{2N}$ is fulfilled. It is important to note that, in contrast with the Hamiltonian in equation~(\ref{eq:Hgoetogue}) for $\alpha=0$, our density matrix has rank one at $t=0$ and therefore it does not belong to the GOE. For nonzero values of $t$, the matrix given in equation~(\ref{crossoverrho}) belongs to the GOE and is the one responsible for the GOE-like statistics at short time. Numerical calculations show this GOE-like spectral statistics at short times. For long times, as the density matrix fills up completely with random variables, the correspondence with the GUE becomes a more accurate description. We will show this behavior relying on numerical calculations in the next section.


\section{Spectral statistics}
\label{sec:specstat}

\begin{figure}
\centering
\includegraphics[width=0.690\textwidth]{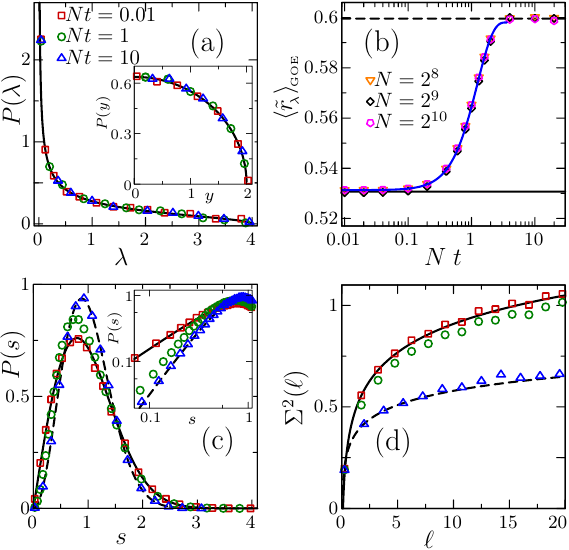}
\caption{(Color online) (a) Spectral level density of the mixed state compared to the Marchenko-Pastur law of equation~(\ref{eq:MP}) in solid line. The inset shows the spectral level density of $y=\sqrt{\lambda}$ compared to the quarter-circle law in black line. (b) $\langle \tilde{r}_{\lambda} \rangle_{\mathrm{GOE}}$ as a function of time rescaled to the system size $N$. The blue curve corresponds to $f(Nt/2\sqrt{3}) - 0.0052$ with $f(\alpha)$ given by equation~(\ref{eq:rtAna1}). (c) Nearest-neighbor level spacings distribution compared to the Wigner-Dyson surmise, equation~(\ref{eq:WD}). A log-log plot is shown in the inset. (d) Level number variance compared  to the GOE and GUE expectations as indicated in equation~(\ref{eq:LNV}). GOE (GUE) predictions are indicated in panels (b)-(d) with solid (dashed) black curves. In (a), (c) and (d) the system size is $N=2^{10}$, and the different times are indicated in the legend of (a). }
\label{fig:RMTfig}
\end{figure}

In this section we study the statistical properties of the mixed state of equation~(\ref{eq:RT}) when the time evolution is generated by an ensemble of Hamiltonians, where each member $H_{l}$ is taken from the GOE, with $l=1, \ldots, \Ne$. For our statistical analysis, we will employ $\Ne=N$, as this is the minimum requirement to obtain a density matrix of the same rank as the dimension $N$. At the end of the section we will show that the main result persists for finite $\Ne>N$. The initial state $| \Psi_{0} \rangle =(1\;0\;\ldots\;0)^{T}$, where $T$ stands for the transpose, is chosen from the canonical basis of the Hamiltonian. For the analysis we consider system sizes of $N=2^{n}$ with $n=8,9$, and 10, and we perform $M$ random realizations of equation~(\ref{eq:RT}) such that the product $N\times M$ remains fixed to $\approx5\times10^{5}$. Also, for each realization we take only $60\%$ of the eigenvalues around the center of the spectrum. For the density matrix as evaluated in equation~(\ref{eq:RT}), we will show that the distribution of the eigenvalues $\lambda_i$, normalized to its standard deviation, follows a Marchenko-Pastur distribution given by
\begin{equation}
\label{eq:MP}
P(\lambda) = \frac{1}{2\pi} \sqrt{\frac{4}{\lambda} - 1} , \quad \lambda \in [0, 4] ,
\end{equation}
which is a characteristic feature of structureless ensembles~\cite{Zyczkowski2011}. For small times this distribution is only reached by all but one of the eigenvalues, the largest one separated from the bulk.

A systematic form of sampling random density matrices is given by taking partial trace of random pure states in a Hilbert space of dimension $N\times N_{\rm A}$, where $N_{\rm A}$ is the dimensionality of the degrees of freedom being traced out~\cite{Zyczkowski2011}. In this case, it has been shown that the eigenvalues of the reduced density matrix follows a Marchenko-Pastur distribution that for $N=N_{\rm A}$ has the form in equation~(\ref{eq:MP}). In our case, we use an ensemble of randomly evolved pure states to generate the mixed state in equation~(\ref{eq:RT}). However, any mixed state can be regarded as the reduced density matrix of a pure state of a larger composite system. Therefore, the eigenvalues of any mixed state can be understood as the Schmidt coefficients squared in the Schmidt decomposition of a particular purification of the density matrix~\cite{Nielsen00}.

In figure~\ref{fig:RMTfig} (a) we display the spectral level density distribution of the mixed state compared to the Marchenko-Pastur law (continuous line)
for small (red-squares), intermediate (green-circles), and large times (blue-triangles), rescaled to the system size $N=2^{10}$. As it can be observed, the same behavior is preserved in time. In the inset of the same panel, we show that the level density follows the quarter-circle law, $P(y) =\frac{1}{\pi} \sqrt{4 - y^{2}}$, in the rescaled variable $y = \sqrt{\lambda}$, characteristic of the so-called Hilbert-Schmidt ensemble of random quantum states~\cite{Sommers2004}.

A further insight about the complexity in the time evolution of the mixed state can be obtained from the average ratio of consecutive eigenvalues spacings, $\langle r \rangle$ or $\langle \tilde{r} \rangle$. These quantities are often used as a measure of the degree of chaoticity of a physical system~\cite{Oganesyan2007,Atas2013a}. We will employ analogue quantities for the eigenvalues of our density matrix. That is, given an ordered set of eigenvalues, $\{ \lambda_{i} \}$, the nearest neighbor spacings are $\Delta_{i} = \lambda_{i+1} - \lambda_{i}$ and the ratio of consecutive eigenvalues spacings, $r_{i}$, and $\tilde{r}_{i}$ are defined as
\begin{equation}
r_{i} = \frac{\Delta_{i}}{\Delta_{i-1}} \quad \mathrm{and} \quad \tilde{r}_{i} = \mathrm{min}\left( r_{i}, \frac{1}{r_{i}} \right).
\end{equation}
Furthermore, the dynamical behavior of the average value $\langle \tilde{r}\rangle$ can be expressed as a function of the parameter $\alpha$, introduced in equation~(\ref{eq:Hgoetogue}), as $\langle \tilde{r}\rangle\equiv\langle \tilde r(\alpha)\rangle=f(\alpha\sqrt{N/3})$, where we have made use of the following auxiliary expression~\cite{Sarkar2020}
\begin{equation}
\begin{aligned}
f(\alpha) =&\frac{4 \left(2+\alpha ^2\right)}{\pi  \left(1-\alpha^2\right)}\arctan\left[\frac{\sqrt{3}(1+\alpha^2)}{2 \alpha }\right]
-\frac{4 \sqrt{3}}{\pi  \left(1-\alpha^2\right)^{3/2}} \arctan\left[\frac{\left(1-\alpha^2\right)^{3/2}}{\alpha (3+\alpha^2)}\right] 
 \\
&-\frac{17+7 \alpha^2}{\pi  \left(1-\alpha^2\right)}\arctan\left(\frac{\alpha }{\sqrt{3}}\right)-\frac{1}{\pi}\arctan\left(\sqrt{3}\, \alpha \right),
\quad 0\le \alpha\le 1.
\end{aligned}
\label{eq:rtAna1}
\end{equation}
This result was derived in the Hamiltonian context and describes the crossover from GOE to GUE in the case of $3\times 3$ matrices. Noting that the transition scales as $\sqrt N$~\cite{Sarkar2020}, here we employ $f(\alpha \sqrt{N/3})$ in order to extrapolate this result to arbitrary values of $N$. In the limit $\alpha\rightarrow0^{+}$ ($\alpha\rightarrow1$) equation~(\ref{eq:rtAna1}) produces $\approx0.5359\;(\approx0.6027)$ for the GOE (GUE), however, for large matrices a more accurate value is~\cite{Atas2013a} $\langle \tilde{r} \rangle^{\mathrm{fit}}_{\mathrm{GOE}} \approx 0.5307$ ($\langle \tilde{r} \rangle^{\mathrm{fit}}_{\mathrm{GUE}} \approx 0.5996$). We will show that the result in equation~(\ref{eq:rtAna1}) also describes the crossover of the ensemble of density matrices introduced in section~\ref{ensemble}. By comparing Eqs.~(\ref{eq:Hgoetogue}) and (\ref{crossoverrho}), one can note that in the case of the density matrix, one requires the replacement $\alpha\rightarrow \sqrt{N}t/2$ in order to obtain an analytical expression for this quantity as $\langle \tilde r_\lambda(\sqrt{N}t/2)\rangle=f(Nt/2\sqrt{3})$. For sake of clarity, we have included a subscript $\lambda$ in order to indicate that this quantity is calculated from the eigenvalues  $\{\lambda_{i}\}$ of the density matrix in equation~(\ref{eq:RT}).

In figure~\ref{fig:RMTfig} (b), we show $\rgoe$ as a function of $Nt$ for system sizes $N=2^{n}$ with $n=8, 9$, and 10, in orange inverted-triangles, black diamonds, and magenta pentagons, respectively. We observe that all the curves collapse into each other, indicating that the results do not change by increasing the dimensionality of the system. Furthermore, finite size effects are removed by regarding a small eigenvalue window of about $60 \%$ around the center of the spectrum. The black horizontal lines correspond to RMT prediction fittings, $\langle \tilde{r} \rangle^{\mathrm{fit}}_{\mathrm{GOE}} \approx 0.5307$ (continuous) and $\langle \tilde{r} \rangle^{\mathrm{fit}}_{\mathrm{GUE}} \approx 0.5996$ (dashed), for the GOE and GUE respectively~\cite{Atas2013a}. For small times, $\rgoe$ agrees with RMT predictions for the GOE. For intermediate times and in accordance with equation~(\ref{eq:guetime}), the eigenvalue statistics departs from GOE and completely reaches GUE at times $t\approx2/N$. This transition is analog to the transition from GOE to GUE that occurs in Hamiltonian systems when time reversal symmetry is broken by a magnetic field. Here, however, the transition does not have to do with the breaking of time-reversal symmetry, but with the number of statistically independent elements in the mixed state, which is $N (N - 1)/2$ for the GOE and $N(N - 1)$ for the GUE. For short times we have $N'= N(N-1)/2$ real elements in equation~(\ref{crossoverrho}). Finally, the blue continuous line corresponds to $f(Nt/2\sqrt{3})-0.0052$, where $f(\alpha)$ is given by equation~(\ref{eq:rtAna1}). The shift $-0.0052$ accounts for the difference between the values $\langle\tilde r\rangle_{\rm GOE}$ and $\langle \tilde r\rangle_{\rm GOE}^{\rm fit}$, as the first one is derived for $3\times 3$ matrices, while the latter is obtained for large values of $N$.

Now, in order to characterize the spectral level density, we use the Wigner-Dyson nearest-neighbor level spacings distribution and the level number variance. On one hand, the Wigner-Dyson distribution characterizes the short-range correlations of the spectrum and is usually used as an indicator of the chaoticity of a given physical system. To a good approximation, it is given by the so-called Wigner surmise~\cite{Mehta2004}
\begin{equation}
\label{eq:WD}
P_{\rm WD}(s)=
\left\{
\begin{array}{lr}
\frac{\pi}{2}s\exp\left(-\frac{\pi}{4}s^2 \right) &\, \, \mathrm{for}\, \, \,  \mathrm{GOE} , \\
\\
\frac{32}{\pi^2}s^2\exp\left(-\frac{4}{\pi}s^2 \right) & \, \, \mathrm{for}\, \, \, \mathrm{GUE} .
\end{array} \right.
\end{equation}
In equation~(\ref{eq:WD}), $s=\Delta_{i}/\langle \Delta\rangle$ being $\langle\Delta\rangle$ the mean level spacing. On the other hand, the long-range level correlations are characterized through the level number variance, or the variance of the number of unfolded eigenvalues in an interval of length $\ell$, namely
\begin{equation}
\label{eq:LNV}
\Sigma^{2}(\ell) \approx
\left\{
\begin{array}{lr}
\frac{2}{\pi^{2}} \Big[ \mathrm{log}(2\pi \ell) + \gamma + 1 - \frac{\pi^{2}}{8}  \Big] & \, \, \mathrm{for}\, \, \, \mathrm{GOE} , \\
\\
\frac{1}{\pi^{2}} \Big[ \mathrm{log}(2\pi \ell) + \gamma + 1  \Big] & \, \, \mathrm{for}\, \, \, \mathrm{GUE} ,
\end{array} \right.
\end{equation}
where $\gamma \approx 0.577\dots$ is the Euler constant. In order to separate the universal from the non-universal behavior in the spectral density, we first perform the unfolding procedure that yields to a spectral level density in which the mean level spacing is equal to unity~\cite{Mehta2004}. Here, the unfolding procedure
is performed by considering the portion of the spectrum to be unfolded in an ordered sequence $\{\lambda_{i}\}$ and then the staircase function $\eta(\lambda)=\sum_{i}\Theta(\lambda - \lambda_{i})$ is fitted by a seventh-degree polynomial function $\bar{\eta}(\lambda)$. Thus the unfolded spectrum is given 
by~\cite{Guhr1998} $\{\bar{\eta}(\lambda_{i})\}$ and the fitting is performed via the least squares method. $\Theta(x)$ denotes the Heaviside function.

In figure~\ref{fig:RMTfig} (c), we show the nearest-neighbor level spacing distribution, $P(s)$, of the mixed state for a system size $N=2^{10}$, for small (red-squares), intermediate (green-circles), and large times (blue-triangles): $Nt = 0.01$, $Nt=1$, and $Nt = 10$, respectively. The black curves correspond to the Wigner-Dyson surmise of equation~(\ref{eq:WD}) for the GOE (continuous) and the GUE (dashed) cases. As it is observed, for small and large times these statistics are in agreement with the Wigner surmise for the GOE and GUE cases, respectively. This is in accordance with the behavior observed in panel (b) of the same figure. Furthermore, for intermediate times the distribution is neither GOE nor GUE, but a distribution in between both statistics. Also, let us notice that for small level spacings, $s$, the degree of level repulsion increases smoothly from linear to quadratic, or from GOE to GUE, as the system evolves in time without the presence of any magnetic field, as usually happens in Hamiltonian systems in RMT. In the inset of the same panel, this transition in a log-log plot is shown.

In figure~\ref{fig:RMTfig} (d), we display the long-range correlations of the spectra, characterized by the level number variance [see equation~(\ref{eq:LNV})], for the three mentioned times. Again, the transition from GOE to GUE is clearly observed.

\begin{figure}
\centering
\includegraphics[width=0.715\textwidth]{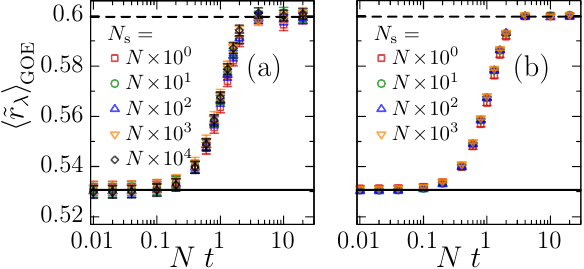}
\caption{(Color online) $\rgoe$ averaged over several ensemble sizes $\Ne$ as function of time for system sizes of $N=2^{6}$ (a) and $2^{8}$ (b). The error bars are the standard error of the mean value.}
\label{fig:Ne_eta}
\end{figure}

Up to this point, we have restricted our analysis to a number of elements in the ensemble equal to the dimension of the system, i.e., $\Ne=N$. Nevertheless, the transition shown in figure~\ref{fig:RMTfig} (b) does not depend on this restriction, as can be inferred from the short-time analysis given in section~\ref{shorttime}. From equation~(\ref{crossoverrho}) one can note that the transition persists even if $\Ne$ grows larger, as $\rgoe$ is evaluated from the ratio of consecutive eigenvalue spacings where the factor $1/\sqrt\Ne$ is cancelled. This can be corroborated in figure~\ref{fig:Ne_eta}, where we plot $\rgoe$ for different values of $N$ and $\Ne$ with $\Ne\ge N$. It is evident that the crossover from GOE to GUE statistics is preserved.  

Before finalizing this section, we  comment
about the strict limit $\Ne\to\infty$ for fixed value of $N$.
In this limit, one can note that equation~(\ref{crossoverrho}) vanishes, as there
is an overall factor $t/\sqrt{2N_s}$ multiplying the  matrices $B$ and $itD\sqrt{N}/2$, where $B$ and $D$ have
finite variances as presented in equation~(\ref{sigmarho}).
The only contribution to the density matrix 
in equation~(\ref{rhosplit}) is given by the 
second order term in time of equation~(\ref{sigmatilde})
as $a_n(t)=Nt^2/2\delta_{n,0}$.
Therefore, one actually obtain an analytical expression for the density matrix that is given by 
$\rho(t)=(1-N t^2/2)|0\rangle\langle 0|+\mathbbm{1}t^2/2$.
It is evident that a crossover to GUE statistics is not possible in this case. However, this only happens in the
strict limit $\Ne\to \infty$ that
corresponds to averaging over the whole orthogonal ensemble. The analysis presented in this section is valid for finite values of $\Ne$. This situation is physically relevant and is motivated in part by limitations of experimental setups such as those dealing with disordered many-body quantum systems where only a finite number of realizations can be performed, see for instance \cite{Lukin2019}.

%


\section{Evolution under interacting spin-1/2 Hamiltonians}
\label{sec:Heisenberg}

This section is devoted to corroborate our previous results based on RMT in a more realistic context. That is, we show that the transition with time to the GUE distribution also occurs when the dynamics of the density matrix in equation~(\ref{eq:PsiofT}) is dictated by the Hamiltonian
\begin{equation}
\label{eq:HamS}
H=\sum_{k=1}^{L} 
\left(S_k^x S_{k+1}^x + S_k^y S_{k+1}^y + S_k^z S_{k+1}^z\right)+\sum_{k=1}^L h_k S_k^z.
\end{equation}
This Hamiltonian describes a one-dimensional system of spin-$1/2$ particles interacting between nearest neighbors, each particle is located in one of the $L$ sites of a chain and subject to an on-site disordered magnetic field. It has become paradigmatic in studies of the so-called many-body localization (MBL) transition/crossover~\cite{Pal2010,Abanin2017,Alet2018}. In Hamiltonian~(\ref{eq:HamS}) $\hbar=1$, $S^{x,y,z}_k$ are spin-$1/2$ operators acting on site $k$ and $h_k$ are random numbers from a flat distribution in the interval $(-h, h)$, with $h$ being the disorder strength. Additionally, periodic boundary conditions are assumed, $S_{L+1}^{x,y,z}=S_1^{x,y,z}$. The Hilbert space dimension is $\mathrm{dim}(\mathcal{H})=2^{L}$, but since the total magnetization in $z$-direction ${\mathcal{S}}^{z}=\sum_{k=1}^LS_{k}^z$ is conserved, the whole space splits in $L+1$ subspaces, each one corresponding to a fixed value of ${\cal{S}}^{z}$ and of size $N=L!/[N_\mathrm{up}!(L-N_\mathrm{up})!]$, with $N_\mathrm{up}$ the number of spins up or excitations contained in each subspace. We concentrate our analysis on two subspaces, namely in the half-filling subspace where $N_\mathrm{up}=L/2$ and in the one-excitation subspace where $N_\mathrm{up}=1$.

For a fair comparison with our GOE results, we ensure that the system described by equation~(\ref{eq:HamS}) is in the chaotic regime. That is, we fix the disorder strength to $h = 0.5$ for the half-filling subspace and for the one-excitation subspace the value of $h$ must be adjusted for each system size as will be specified in section~\ref{ss:oe}. For these cases, the spectrum of equation~(\ref{eq:HamS}) reproduces the main features of time-reversal symmetric systems of the GOE ensembles meaning that its energy levels are correlated in a similar way as the eigenvalues of GOE matrices~\cite{Torres2019,sorathia2012closed,torres2019level}. Note that the case of the one-excitation subspace could be considered analogous to a single-particle model, just as the finite size one-dimensional Anderson model for which the energy spectrum can follow GOE statistics for certain disorder strengths~\cite{sorathia2012closed,torres2019level}. Here, however, it is worth stressing that the crossover from GOE-like to GUE-like statistics in the spectrum of the density matrix arises independently of the regime, chaotic (thermalizing, ergodic), integrable or localized, of the Hamiltonian dictating the time evolution as long as the Hamiltonian is real and contains an enough number of statistically independent random elements, as we will see below. Previously, we mentioned that our choice of initial state in the RMT case was equivalent to a random real state. In this section we start from a random real state, as the Hamiltonian in equation~(\ref{eq:HamS}) is not invariant under orthogonal transformations. Therefore, in both cases the initial state is a pure state whose components are real random variables following a Gaussian distribution with zero mean value and possessing, before normalization, unit variance.


\subsection{Half-filling subspace}
\label{ss:hf}

The half-filling (HF) subspace corresponds to ${\cal{S}}^z=0$ which leads to $N=L!/(L/2)!^2$. For our analysis we set $L=10,\;12$ and $14$; therefore, $N=252,\;924$ and $3432$, respectively. As stated in section~\ref{ensemble}, we perform $M=1984,\;541$ and $146$ realizations in equation~(\ref{eq:RT}) so that for the spectral statistics we count with almost $5\times10^{5}$ eigenvalues on each system size, respectively.
In order to ensure a GOE statistics of the energy levels of the  Hamiltonian we choose $h=0.5$ in the 
analysis of this HF subspace~\cite{Torres2019}.

\begin{figure}
\centering
\includegraphics[width=0.710\textwidth]{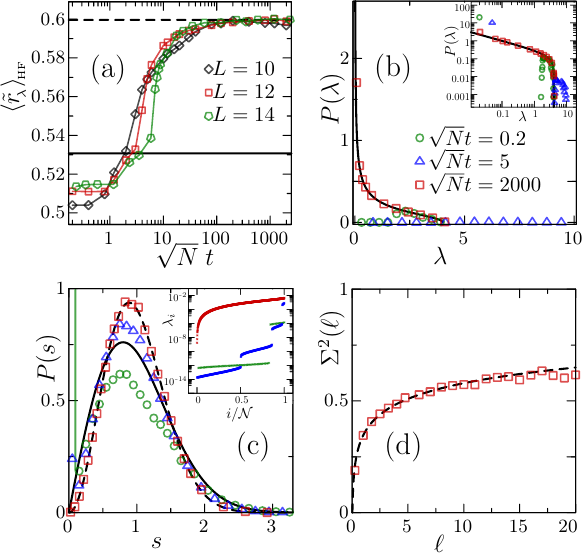}
\caption{(Color online) Spectral statistics of the density matrix 
ensemble~(\ref{eq:RT}) when the time evolution is dictated by an ensemble of 
Hamiltonians taken from the spin-1/2 model~(\ref{eq:HamS}) in the HF subspace with 
$h=0.5$. 
(a) $\langle \tilde{r}_\lambda \rangle_{\mathrm{HF}}$ as a function of time 
rescaled by $\sqrt{N}$ being $N=L!/(L/2)!^2$ the size of the HF subspace for $L$ spins. 
(b) Spectral level density $P(\lambda)$ compared to the Marchenko-Pastur law (black solid curve) given 
by equation~(\ref{eq:MP}).
A log-log version is presented in the inset.
(c) Nearest-neighbor level spacings distribution compared to the Wigner-Dyson 
surmise, equation~(\ref{eq:WD}).  
A semi-log plot of the eigenvalues of one single realization of the
density matrix is presented in the inset where $\mathcal{N}$ is the number of 
available eigenvalues in each case.
(d) Level number variance.
In panels (b)-(d) the system size is $L=12$ ($N=924$)
and the 
different times are indicated in the legend of (b).
In panels (a), (c) and (d) the black solid (dashed) line corresponds to the 
GOE (GUE) prediction.}
\label{fig:XXZHF}
\end{figure}

It is worth mentioning that when simulating the dynamics in the half-filling subspace we realized that for short times, $t\ll1$, many eigenvalues were smaller than $10^{-15}$. Therefore, it was necessary to impose a restriction over the eigenvalues to be considered in our statistical analysis, thus avoiding possible numerical machine precision errors. We consider the whole spectrum in descending order $\lambda_{i}>\lambda_{i+1}$, then we take a subset of that spectrum $\left\{\lambda_{i} \right\}_{i=1}^{\mathcal{N}\leq N}$, such that $\sum_{i=1}^{\mathcal{N}}\lambda_{i}=1-10^{-12}$. In this way, we found that all discarded eigenvalues were smaller than $10^{-15}$ and that for short times $\mathcal{N}\ll N$.  For $t\gtrsim 0.1$ we found $\mathcal{N}\approx N$. This behavior can be understood as at short times the state remains almost pure, while at large times the mixture of the state is increased. In the following analysis we also do not consider the largest eigenvalue $\lambda_{1}$, i.e., we work with the subset $\left\{\lambda_{i} \right\}_{i=2}^{\mathcal{N}\leq N}$.

Figure~\ref{fig:XXZHF} (a) depicts $\rhf$ in the HF subspace as a function of the rescaled time $\sqrt{N}t$ for several system sizes $L=10$ (black diamonds), $12$ (red squares) and $14$ (green pentagons). The first thing to notice is that the universal transition to GUE statistics also occurs in this case regardless the system size, i.e., for large times $\rhf$ $\rightarrow$ $\langle \tilde{r} \rangle^{\mathrm{fit}}_{\mathrm{GUE}}$ (black dashed line); however, in contrast with the case when the dynamics is dictated by GOE matrices, the starting value of $\langle \tilde{r}_\lambda \rangle_{\mathrm{HF}}$ does not coincide with $\langle \tilde{r} \rangle^{\mathrm{fit}}_{\mathrm{GOE}}$ (black solid line). Specifically, for $\sqrt{N}t\lesssim 1$ we observe $\rhf \lesssim 0.51 < \langle \tilde{r} \rangle^{\mathrm{fit}}_{\mathrm{GOE}}$, meaning that the correlations between eigenvalues of the density matrix evolved with Hamiltonians with the form given by equation~(\ref{eq:HamS}) are weaker than those from the density matrix evolved with GOE matrices, $\langle \tilde{r} \rangle^{\mathrm{fit}}_{\mathrm{GOE}}$. After the initial plateau, $\rhf$ shows a transient increasing within the time interval $1\lesssim \sqrt{N} t\lesssim 100$. Finally, when the time is large enough, $\sqrt{N} t\gtrsim 100$, the statistics reaches and remains in the $\langle \tilde{r} \rangle^{\mathrm{fit}}_{\mathrm{GUE}}$ value. 
We have presented results for three different values of the number of spins, namely $L=10,12$ and $14$, where one can note that the transition approximately occurs at the same rescaled time $\sqrt{N}t$.

The lack of agreement between $\rhf$ and $\langle \tilde{r} \rangle^{\mathrm{fit}}_{\mathrm{GOE}}$ at the short-time scale is explained as follows. Since at half-filling the Hamiltonian~(\ref{eq:HamS}) contains only $L$ statistically independent random variables, a number which is in general much smaller than $N$,  the CLT cannot be applied to equations~(\ref{matricesrhoa}) and~(\ref{matricesrhob}); thus, the elements of matrices $B$ and $D$ in equation~(\ref{crossoverrho}) cannot be taken as uncorrelated Gaussian random variables. In order to reinforce our previous arguments we will show in the next subsection that an enough number of independent random variables in Hamiltonian~(\ref{eq:HamS}) leads to initial eigenvalues statistics that agree with $\langle \tilde{r} \rangle^{\mathrm{fit}}_{\mathrm{GOE}}$. Before that, let us complete our analysis by studying the distribution $P(\lambda)$ for the eigenvalues $\lambda_{i}$ of the density matrix ensemble in the HF subspace. 

In figure~\ref{fig:XXZHF} (b) we display the distribution $P(\lambda)$ for $L=12$ at times $\sqrt{N}t=0.2$ (green circles),  
$\sqrt{N}t=5$ (blue triangles), and $\sqrt{N}t=2000$ (red squares). 
The agreement with the Marchenko-Pastur distribution given in equation~(\ref{eq:MP}) (black line) is only observed for times $\sqrt{N} t\gtrsim2000$ and where, as already discussed, $\rhf$ coincides with the GUE statistics. For short times (green circles) and intermediate times (blue triangles) $P(\lambda)$ is far from the Marchenko-Pastur distribution. Therefore, we can ensure that only at large times the density matrix generated according to equation~(\ref{eq:RT}) and evolved with Hamiltonian equation~(\ref{eq:HamS}) in the HF subspace can be considered as a genuine random density matrix from a Hilbert-Schmidt ensemble. This marks another important difference with the density matrix evolved with GOE matrices, for which the distribution $P(\lambda)$ agrees very well with the Marchenko-Pastur distribution from short to large times [see figure~\ref{fig:RMTfig} (a)]. 

The nearest-neighbor level spacing distribution 
is presented in figure \ref{fig:XXZHF} (c)
for $L=12$ and for the same three values of time as indicated in the legend of panel (b). 
For a short time (green circles), the distribution displays a behavior which is below of the 
GOE statistics (black solid line). 
For large times, GUE statistics (black dashed line) is reached as depicted with the
red squares. Intermediate time is presented with the blue triangles. 
Those results are in accordance with the predicted by $\rhf$ in 
figure~\ref{fig:XXZHF} (a).
In the inset of figure~\ref{fig:XXZHF} (c), we present the 
eigenvalues of the density matrix of a single realization with the same color 
code as in the main panel. 
One can note that for short times, the spectrum displays gaps, a situation that 
complicates the unfolding procedure as described in 
section~\ref{sec:specstat}. 
However, we can overcome that difficulty by employing another unfolding method which
consists on splitting the total spectrum  in several subsets of eigenvalues
and then divide each eigenvalue in the subset by its corresponding mean level spacing. 
The price to pay is that such unfolding only captures the short range correlations 
and therefore it is not possible to calculate the level number variance.
For this reason, we only present the level number variance for large times in 
figure \ref{fig:XXZHF} (d), where the gaps in the spectrum are absent as noted from 
the inset in figure~\ref{fig:XXZHF} (c).
It can be noted that  for large times the long range correlations corresponds to that of
the GUE statistics.


\subsection{One excitation subspace}
\label{ss:oe}

In this section we repeat the analysis done in subsection~\ref{ss:hf} for a density matrix which is evolved under Hamiltonians with the form given in equation~(\ref{eq:HamS}), but now in the subspace where only one excitation (OE) is present in the chain. The dimension of this subspace is $N=L$, a convenient fact for our purposes because in this case the number of statistically independent random variables in the Hamiltonian equals the dimension of the subspace. We set $L=500,\;924,$ and 1500, and we perform $M=1000, \;541,\; 333$ random realization, respectively. The disorder 
strength in equation~(\ref{eq:HamS}) is set to $h=0.15, \;0.1,$ and 
0.09 in order to ensure a GOE statistics of the energy levels
of the Hamiltonian 
for each system size aforementioned~\cite{sorathia2012closed,torres2019level}.
The initial state is the same one used for the analysis in the HF subspace.

\begin{figure}
\centering
\includegraphics[width=0.710\textwidth]{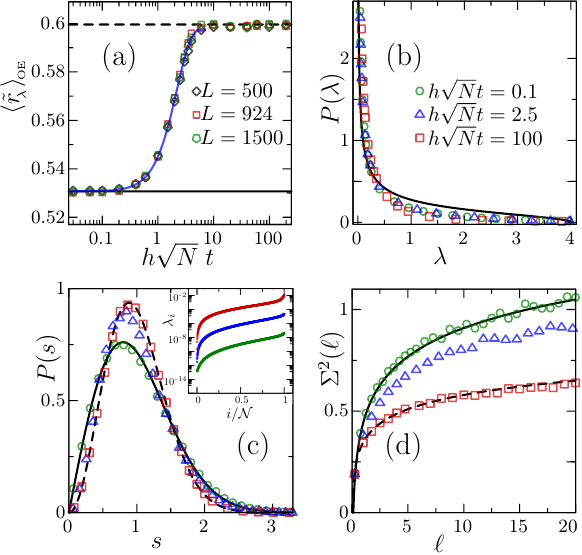}
\caption{ (Color online) Spectral statistics of the density matrix ensemble~(\ref{eq:RT}) when the time evolution is dictated by an ensemble of Hamiltonians taken from the spin-1/2 model ~(\ref{eq:HamS}) in the OE subspace. 
(a) $\langle \tilde{r}_\lambda \rangle_{\mathrm{OE}}$ as a function of time 
rescaled by $h\sqrt{N}$ being $h$ the disorder strength and $N=L$ the
size of the OE subspace. Black  solid (dashed) line is the $\langle \tilde{r}\rangle^{\mathrm{fit}}_{\mathrm{GOE}}$ ($\langle 
\tilde{r}\rangle^{\mathrm{fit}}_{\mathrm{GUE}}$). 
The blue solid curve corresponds to $f(at)- 0.0052$ with $f(\alpha)$ given by equation~(\ref{eq:rtAna1}) and $a=0.4818\pm0.0004$. 
(b) Spectral level density $P(\lambda)$ compared to the Marchenko-Pastur law 
given by equation~(\ref{eq:MP}) (black curve). 
(c) Nearest-neighbor level spacings distribution compared to the Wigner-Dyson
surmise, equation~(\ref{eq:WD}). A semi-log plot of the eigenvalues of one
single realization of the density matrix is presented in the inset where
$\mathcal{N}$ is the number of available eigenvalues in each case.
(d) Level number variance.
In panel (a) $h=0.15,0.1$ and $0.09$ for $L=500,924$ and $1500$, respectively.
In panels (b)-(d) only the case $L=924$ ($N=924$) with $h=0.1$ is considered 
and the times are indicated in the legend of (b).
In panels (a), (c) and (d) the black solid (dashed) are the GOE (GUE) 
predictions.}
\label{fig:XXZOE}
\end{figure}

Figure~\ref{fig:XXZOE} (a) shows our numerical results for the dynamical statistical behavior of $\roe$ for $L=500$ (black diamonds), 924 (red squares), and
1500 (green pentagons). We observe that $\roe$ agrees for short times with $\langle\tilde{r}\rangle^{\mathrm{fit}}_{\mathrm{GOE}}$ (black solid line), while for large times there is an agreement with $\langle\tilde{r}\rangle^{\mathrm{fit}}_{\mathrm{GUE}}$ (black dashed line) regardless of the system size. This is a remarkable qualitative correspondence with the behavior of $\langle\tilde{r}_\lambda\rangle_{\mathrm{GOE}}$ [see figure~\ref{fig:RMTfig} (b)]. Furthermore, $\roe$ is well described by equation~(\ref{eq:rtAna1}) via $f(at)-0.0052$ (blue solid curve), being $a=0.4818\pm0.0004$ the best fitting parameter in the validity range $0\leq h\sqrt{N} t\lesssim 6$. The curve fittings are performed using a nonlinear least-squares fitting.
Notice that we have
identified that the rescaling in time in this case
is given by $h\sqrt{N} t$. The presence of the disorder strength in the rescaling prefactor is in line with
previous findings in the context of tight-binding models for a single particle propagating in one-dimensional disordered media, see for instance \cite{sorathia2012closed,torres2019level}.

For completeness, we show in figure~\ref{fig:XXZOE} (b) the distribution $P(\lambda)$ for the eigenvalues of the density matrix, but now evolved in the OE subspace for 
$h\sqrt{N}t=0.1$ (green circles), $2.5$ (blue triangles), and $100$ (red squares), only the case $N=924\;(h=0.1)$ is considered. In any case the Marchenko-Pastur distribution (black curve)
is not reached. At first glance this could be surprising, specially because at $h\sqrt{N}t=100$ we have that $\langle\tilde{r}_\lambda\rangle_{\mathrm{OE}}$ already coincides with $\langle\tilde{r}\rangle^{\mathrm{fit}}_{\mathrm{GUE}}$, however, it is known that Hamiltonian~\eqref{eq:HamS} in the subspace of a single excitation does not depict many-body chaotic features, see for instance Ref.~\cite{zisling2021many}. Furthermore, differences between global properties of many-body systems in subspaces with few and several excitations are well known ~\cite{schiulaz2018few}. 

In addition to the previous discussion, let us note that for the corresponding disorder strength $h$ used in each case, HF ($h=0.5$) and OE ($h=0.1$) subspaces, the level spacing distribution of Hamiltonian~(\ref{eq:HamS}) is GOE-like; however, there is an essential distinction between both cases, namely for the HF subspace an infinitesimal disorder strength could bring the system to a chaotic regime in the thermodynamic limit ($L\to\infty$)~\cite{santos2020speck}, meanwhile for the OE subspace in the same limit, an infinitesimal disorder strength will induce single-particle-like localization and the level spacing distribution will be the corresponding to uncorrelated random variables from a Poisson process~\cite{sorathia2012closed,torres2019level}. Interestingly, this fact could be indicating that the distribution $P(\lambda)$ of eigenvalues of the random density matrix~(\ref{eq:RT}) is more sensitive for the characterization of the chaotic nature of a given finite system than the level spacing distribution of the associated Hamiltonian. But this last point needs further studies.

In figure \ref{fig:XXZOE} (c), we present
the nearest-neighbor level spacings distribution for the same times as
indicated in the legend of panel (b) and with $L=924$, where again
one can note correspondence of the statistics
to GOE and GUE for short and large time respectively.
As can be observed from the inset, there are no gaps in the spectrum as in the case of the HF subspace.
Therefore, the unfolding can be performed as discussed
in section~\ref{sec:specstat} allowing us to calculate the level number variance
which is shown in figure \ref{fig:XXZOE} (d) corroborating the correspondence
to the GOE and GUE statistics for short and long times
respectively.

Above all, in this subsection we have validated that the number of statistically independent real random variables needed in order to have a crossover from GOE to the universal GUE in the statistics of the unfolded eigenvalues of the density matrix generated according to equation~(\ref{eq:RT}) should be at least ${\cal{O}}(N)$. This conjectured lower bound is based in the fact that for the spin $XXX$ model in the HF subspace contains only $L$ independent random numbers, while in the OE subspace the model contains $N$ independent random numbers. An amount larger than $N$ does not change the picture, as it is apparent when the evolution is dictated by GOE matrices, for which the number of independent random numbers scales as $N^2$.


\section{Conclusions}
\label{conclusions}

In this paper, we have studied the time dynamics of random density matrices generated by evolving a pure state using two different ensembles of random Hamiltonians. In the first case, we consider an ensemble of Hamiltonians that belongs to the GOE and show that the resulting ensemble of mixed states undergoes a crossover from the GOE to the GUE in their spectral statistics. This crossover occurs in a time scale that depends on the system size and is explained in terms of the elements of the density matrix, which are real or complex uncorrelated Gaussian random variables for short and large times, respectively. 

In the second case, we showed that the GOE to GUE crossover can also be present in more realistic contexts in the eigenvalues statistics of the density matrix. Here, the time evolution of the ensemble of mixed states is  dictated by an ensemble of spin-1/2 Hamiltonians including on-site disorder and displaying GOE statistics. Particularly, two subsectors of the Hamiltonians are considered: one of many-body nature, the hall-filling subspace, and another one with single-particle nature, the one-excitation subspace. For these cases the spectral statistics of the generated mixed state shows, at short times, a strong dependence on the number of statistically independent random variables present in the Hamiltonians. Furthermore, at large times, even though both ensembles of mixed states reach their final universal form, marked by GUE statistics, only mixed states generated by the ensemble of spin-1/2 Hamiltonians in the half-filling subspace lead to genuine random density matrices belonging to the Hilbert-Schmidt ensemble, whose level density follows a Marchenko-Pastur distribution.
 
In general such a crossover is not always present. For instance, using complex random Hamiltonians to generate the evolution, the eigenvalues statistics of the
density matrix will resemble the  GUE for all times. In contrast, using antisymmetric Hamiltonians will lead to a constant GOE behavior. We emphasize, however, that for real Hamiltonians with random components the crossover to GUE statistics studied here marks the onset of universality of dynamically generated random density matrices. 

It would be interesting to study further statistical properties of random density matrices, such as the time dependence of the eigenstates structure, together with the distribution of their components. Additional studies on the eigenvalues distribution could assist in chaos detection in different Hamiltonian scenarios.


\section*{Acknowledgments}

E.J.T.-H., and J.M.T. are grateful for financial support from Proyectos VIEP 2021, BUAP, project number 100524481-VIEP2021. J.M.T. also acknowledges support from the project number 100527172-VIEP2021. A.M.M.-A. acknowledges financial support from CONACyT. M.C.-N. thanks financial support from CONACyT through the grant ``Ciencia de Frontera 2019'', No. 10872. Also, M.C.-N acknowledges the facilities provided by the ``Centro de An\'alisis de Datos y Superc\'omputo (CADS)'' from the University of Guadalajara through of the Leo-Atrox Supercomputer.





\end{document}